VORTEX DYNAMICS CONTROLLED BY LOCAL SUPERCONDUCTING ENHANCEMENT.

V Rollano[1], A Gomez[2], A Muñoz-Noval[3], J del Valle[4], M Menghini[1], M C de Ory[1], J L Prieto[5], E Navarro[3], E M Gonzalez[1,3] and J L Vicent[1,3]

[1]IMDEA-Nanociencia, Cantoblanco, E-28049 Madrid, Spain

[2]Centro de Astrobiología (CSIC-INTA), Torrejón de Ardoz, E-28850 Madrid, Spain

[3]Departamento Fisica de Materiales, Universidad Complutense, E-28040 Madrid, Spain

[4] Department of Physics, Center of Advance Nanoscience, University of California-San Diego, 9500 Gilman Dr, La Jolla, CA 92093, USA

[5]Instituto de Sistemas Optoelectronicos y Microtecnologia, Universidad Politecnica Madrid, E-28040 Madrid, Spain

**Abstract.** A controlled local enhancement of superconductivity yields unexpected modifications in the vortex dynamics. This local enhancement has been achieved by designing an array of superconducting Nb nanostructures embedded in a V superconducting film. The most remarkable findings are: i) vanishing of the main commensurability effect between the vortex lattice and the array unit cell, ii) hysteretic behavior in the vortex dynamics, iii) broadening of the vortex liquid phase and iv) strong softening of the vortex lattice. These effects can be controlled and they can be quenched by reducing the Nb array superconducting performance applying an in-plane magnetic field. These results can be explained by taking into account the repulsive potential landscape created by the superconducting Nb nanostructures on which vortices move.

**Introduction.**

Long time ago Anderson set the focus on the behavior of superconductors at the nanoscale. He explored at which nanomaterial sizes the superconductivity will actually cease [1]. Since then, the current development of nanofabrication techniques has opened a fruitful scenario in this field. Nowadays, mesoscopic superconductivity is a well-established field with very impressive achievements. We can quote Cooper pair box, related to charge qubit in quantum computing [2], superconducting vortex pattern, related to the symmetry imposed by the shape of the nano-superconductor [3], suppression of superconductivity in ultrathin nanowires related to phase slips [4] and so on.

Our main aim is not to study nanosized superconductors in themselves, but to investigate the effect of a distribution of nanosized superconductors in contact with a different plain superconductor. In particular, we investigate the effect of the local enhancement of superconductivity in the mixed state behavior of the plain superconductor. Hence, in the present work, we have engineered an array of superconducting nanodots embedded in a superconducting film whose critical temperature is slightly lower than the array critical temperature with characteristic superconducting lengths being similar in both superconductors. In the literature some works can be found which are focused on this type of hybrid structures, we can mention the study of the crossovers from pinning enhancement to superconducting wire network [5]; and from pinning to antipinning landscapes [6]. On the other hand, using the non-linear Ginzurg-Landau theory and Bitter decoration, Berdiyorov et al. [7] have studied vortex configurations due to superconducting pillars in superconducting films. In all of these studies the interplays among different length scales are crucial. In our work, the dimensions of the arrays and nanodots are chosen to prevent unwanted crossovers to different superconducting regimes as happens in the aforementioned works. In the present work, we show that local enhancement of superconductivity allows modifying mixed state effects in hybrid systems made of two superconductors; the most relevant ones are quenched of the main commensurability effect, softening of pinning forces, broadening of the vortex liquid phase and finally, the emergence of hysteresis effects in the vortex dynamics.

The paper is organized as follows: after a description of the fabrication, characterization and experimental techniques, the results and discussion are presented in two sections: 1. Commensurability effects between the vortex lattice and the array unit cell; 2. Temperature dependence of the vortex dynamics. Finally, a summary section closes the paper.

**Experimental.**

Two hybrid systems have been fabricated on Si substrates by electron beam lithography, sputtering and etching techniques. They consist of equilateral ($l = 612$ nm) nanotriangles embedded in a superconducting V thin film of 100 nm. In the main sample (MS in the following), the nanostructures are made of 40 nm superconducting Nb whereas in the witness sample (WS in the following), they are made of non-superconducting material, in this case, 40 nm of Cu. This WS sample plays a framework role for our study.

We have chosen Nb nanodots of triangular shape, since they have the same symmetry as Abrikosov vortex lattice. Therefore, the triangles can host vortices without distortions. For example, giant vortices, which can exist for instance in mesoscopic superconducting disks [8], are precluded in our study. The array of nanotriangles is shown in figure 1, and figure 1 inset shows sketches of the WS and MS samples.

The samples can be considered comprising two triangles, oriented up and down, with roughly the same dimensions. The superconducting critical temperature of the Nb array is 4.84 K, measured using a SQUID magnetometer. The critical temperature of the WS and MS hybrids are 4.38 K and 4.25 K respectively, measured by transport technique. We have estimated the coherence lengths, $\xi(T)$, measuring the upper critical fields as usual. For the V film ($T_c$ = 4.58 K) with thickness 100 nm, we obtain $\xi(0)$ = 9.3 nm and for the Nb film ($T_c$ = 5.9 K) with thickness 40 nm, we obtain $\xi(0)$ = 8.8 nm. Eight terminals crossed-shape bridge is patterned for measuring magnetotransport properties. These measurements are taken using a commercial He cryostat with a 90 kOe superconducting solenoid, a rotatable sample holder that allows varying the applied field direction in situ, and a variable temperature insert. The transport measurements are taken by the usual four probe dc technique. (I, V) characteristic curves are also measured, critical currents are obtained by using a voltage criterion of 20 µV/cm corresponding to 0.1 µV in the sample. More experimental details can be found in [9].

**Results and discussion.**

*Vortices on the move: Commensurability effects.*

Arrays of defects inside a superconductor are powerful tool to probe and modify vortex dynamics. These arrays can be made of magnetic [10] or non-magnetic dots [11], and holes (antidots) [12] or blind holes (blind antidots) [13]. These systems have yielded a flood of results and relevant effects have been found as, for example, commensurability effects [10-13], reconfiguration of the vortex lattice [14], channeling effects [15,16], ratchet effect [9,17] and so on (see for instance the review [18] and references therein).

First, we study the influence of the Nb array on commensurability effects between the vortex lattice and the "defect" unit cell, in our case an array of superconducting nanodots. These effects generate equally spaced resistivity minima at the matching fields, when the vortex density is an integer number of the density of defects. For example, the first matching field corresponds to the magnetic field where the density of vortices equals that of pinning centers. At these matching fields, the vortex lattice motion slows down and therefore, minima in the resistance (maxima in the critical current) are obtained. The magnetotransport data of our samples are plotted in figure 2 (a). The main result is that first minimum is absent in the MS sample, while in WS sample the main minimum appears. The MS sample result is unexpected at first sight, since the origin of the first minimum is directly related to the geometry of the array unit cell. A double check of the lack of the first minimum can be achieved by means of critical currents vs applied magnetic fields measurements. Figure 2(b) shows clearly that the critical current maximum is absent at the first matching field. It is worth noting that in MS sample, for applied magnetic field up to the third matching field, the minima are sharp and well-defined as usual. Beyond this field, the magnetoresistance data show a structure with shallow and not-well-defined minima. From the comparison of the experimental results of WS and MS samples, we can determine that the origin of this anomalous behavior is related to the superconducting character of the periodic potentials. Usually, commensurability effects are generated by ordered array of nanostructures which produce a local suppression of the superconductivity, generating attractive potentials for the vortices as in the WS sample. On the contrary, in the MS sample the ordered potential is originated by Nb nanotriangles with critical temperature slightly higher than the critical temperature of the V film. This means that near the MS critical temperature $T_{c0}$, the Nb nano-islands expel the vortices. They act in the same way than antipinning centers, creating a repulsive potential, due to a local enhancement of superconductivity, that interact with the vortex lattice. So, vortices move without probing the ordered array and, therefore, there is not commensurability between the vortex lattice and the superconducting array.

The situation changes when the vortex density is increased. As shown in figure 2, commensurability effects show up as minima in the resistance and maxima in the critical current. This is due to caging effects induced in the interstitial vortex lattice by the Nb nanotriangles array, in the same way that was reported in [19]. This is confirmed by adding a third vortex per unit cell, that enhances the caging effect and gives rise to a deeper resistance minimum and larger critical current maximum (see figures 2). Beyond the third matching field, the commensurability effects diminish and smooth out. To figure out this finding, we have to compare the number of interstitial vortices with the so-called filling factor. In general, this factor gives a rough estimation of the number of vortices that can be fitted in a pinning site and depends on the ratio between the dimension of the defects and the superconducting coherence length [20]. In our case, we calculate the filling factor as an estimation of the number of vortices that can be caged in the interstitial area and this number turns out to be 3 at T = 0.97 $T_{c0}$. Therefore, the fourth vortex per unit cell exceeds the filling factor, precluding the commensurability effects. These magnetoresistance results indicate that the synchronized vortex lattice for fields larger than the third matching field is shaped with two types of vortices, interstitial vortices and vortices probing the Nb dots. From these results, we can conclude that commensurability effects can be obtained by repulsive potentials created by local enhancement of superconductivity.

Next, in order to confirm the significance of the superconducting Nb nanodots we seek how to modify their role in the commensurability effects. A way to change the Nb superconducting state is applying in-plane magnetic fields, and a straightforward method for this is rotating the sample in an applied magnetic field. Figure 3 (a) shows the magnetoresistance curves (MS sample at 0.98 $T_{c0}$) at different angles, θ, between the magnetic field and the direction normal to the sample surface. As shown in figure

3 (b) the distance between consecutive resistance minima scales with $1/\cos(\theta)$, which is in agreement with the results of Martin et al. [10].

These authors found that only the perpendicular component of the magnetic field is relevant for the commensurability effects. In our case, the most relevant outcome is that the first minimum emerges when $\theta$ increases beyond $\theta = 50º$, see star symbols in figure 3(a). This indicates that the interstitial sites are not energetically favorable when the magnetic field is tilted beyond this angle and usual commensurability for the first matching field arises. In order to understand this effect, it has to be considered that the potential landscape created by the Nb nanotriangles emerges from both the repulsive potential created by the superconducting character of the nanotriangles, and the attractive one created by the periodic corrugation [21]. Therefore, when the parallel component ($H_\parallel$) of the applied magnetic field increases, the superconducting performance of the Nb nanotriangles is diminished and the antipinning potential is smoothed. Consequently, the origin of the matching effect at the first matching field for $\theta > 50º$ is the attractive potential induced by the periodic roughness of the sample that leads to the usual commensurability effect. Therefore, for large enough in plane magnetic fields, the Nb nanotriangles become potential wells energetically more favorable for the vortices than the interstitial sites.

*Vortices on the move: Temperature effects.*

The competition between the intrinsic random defects and the artificially induced periodic defects governs the vortex dynamics [22, 23]. Commensurability effects exist in narrow temperature windows close to the critical temperature, since reducing the temperature, the pinning by the periodic array becomes weaker than the pinning by random defects. Regarding driving currents, the critical currents settle the limit to move the vortices.

In this section, the vortex dynamics temperature dependence of these effects is investigated. The most remarkable results are shown in figure 4. We observe that decreasing the temperature the minima vanish as expected. In addition, an unexpected feature develops: decreasing the temperature and applying magnetic fields above the third matching field, magnetoresistance curves show hysteresis, see panels 4 (b) and 4(c). The usual behavior (without magnetic hysteresis) occurs in the WS sample, see panel 4 (d) and inset. In the hysteretic region, increasing the applied field leads to higher dissipations (red curves in figure 4) than decreasing the applied field (blue curves in figure 4). This can be explained by means of surface barriers which are different for the entrance and exit of vortices into the superconducting nanotriangles [24 - 27]. This behavior enhances at low temperatures, even when the commensurability effects eventually disappear at low temperatures, see figure 4 (c). According to the discussion in the previous lines, these features are a confirmation of the crucial role played by the Nb nanodots in the vortex dynamics. Increasing or decreasing the applied magnetic field means increasing or decreasing the number of vortices in the sample. The hysteresis is only observed when the number of vortices is higher than three vortices per unit cell which corresponds to the filling factor. When the density of vortices is above the third matching field, interstitial vortices (non-hysteresis) and vortices which probe the Nb dots (hysteresis) coexist in the sample. These results are in contrast to the ones reported by He et al. [28] where the magnetotransport hysteresis in multi-connected superconducting islands is attributed to interstitial vortices solely.

A further proof of the crucial role played for the Nb triangles on the vortex dynamics is to explore the angular dependence of the magnetoresistance hysteresis. As we discussed few lines before, tilting the applied magnetic field allows applying in-plane magnetic field on the Nb triangles and therefore, depressing their superconducting properties. Taking into account this fact, we expect that the hysteresis fades away when the Nb array starts smoothing its superconductivity performance. Figure 5 shows magnetoresistance taken at different tilted angles. The hysteresis disappears for angles higher than θ = 50º.

Finally, the (H, T) diagram in the MS sample is studied in comparison with the standard WS sample. The behavior of (I, V) characteristic curves is the ideal experimental tool to explore the (H, T) diagram. Figures 6 shows the experimental (I, V) curves taken in MS and WS samples. In both cases, we observe that increasing the temperature the vortex behavior evolves continuously from a sharp depinning characteristic curve to almost linear (ohmic) characteristic curve. This behavior is the fingerprint of a transition, in (H, T) phase diagram, from glassy solid to a liquid. Liquid and solid vortex matters arose with high temperature superconductivity (HTS); see for example the review of Blatter et al. [29]. Actually, these vortex states are quite general and vortex solid/vortex liquid transition has been observed in very different superconductors, for example: pnictides [30], organic superconductors [31] and p-wave ferromagnetic superconductor [32]. These two vortex matter regimes called the attention of many researcher and a plethora of new features have been identified in the HTS field, for example Bragg, Bose, and splayed glasses, disentangled, entangled liquids and so on [33 - 41]. From (I, V) characteristic curves the liquid- glassy solid transition temperature ($T_g$) can be found. To obtain $T_g$ we have followed the method and analysis of Strachan et al. [42]. These authors analyzed and discussed the usual way to find the transition temperature [43]. That is based on using scaling analysis of the (I, V) characteristic curves. Strachan et al concluded that the standard approach is not correct; since, using scaling analysis to study the transition, several different critical temperatures can be obtained. They proposed a careful and unambiguous method to determine the critical temperature based in the drastic critical changes in the (I, V) curve concavities.

Following this approach, we obtain the transition temperature $T_g$ by analyzing the derivatives of log(V)-log(I) curves for both samples, see figure 7. Above the transition temperature $T_g$, a maximum appears, which implies low current ohmic tails characteristic of the liquid phase (red curves). Below this temperature, this maximum disappears while maintaining the negative concavity in the (I, V) curves, characteristic of the glassy behavior (blue curves). We have to stress that the MS and WS characteristic curves show the same trends. However, the local enhancement of superconductivity in sample MS has remarkable consequences on the (H, T) diagrams, as we show in the following lines. From analysis in figure 7 panels (a) and (b), $T_g$ is obtained for the MS sample at two different applied magnetic fields. For the third matching field, that corresponds to vortices which do not probe the Nb dots (interstitial vortices only and sharp and well-defined minima), $T_g = 3.95$ K ± 20 mK. For the seventh matching field, that corresponds to vortices probing the Nb dots (shallow and not so well-defined minima), $T_g = 3.93$ K ± 20 mK is obtained. Figure 7 panels (c) and (d) show the transition temperatures ($T_g$) in the same experimental conditions in sample WS. From these graphs, we obtain $T_g$ = 4.25 K ± 20 mK. In comparison with the WS sample, the MS sample shows a broadening of the vortex liquid phase. The liquid region is enlarged roughly from 0.97 $T_{c0}$ (WS sample) to 0.93 $T_{c0}$ (MS sample). In summary, the local enhancement of the superconductivity in the MS sample produces a clear softening of the vortex lattice, which is confirmed with the decrease of the pinning force, $F_c = J_c B$, shown in figure 8, $J_c$ and B being the critical current density and the magnetic field respectively.

Finally, we have to underline that, in the MS sample, the crossover to the vortex solid state activates the magnetoresistance hysteresis, which is absent in the vortex liquid region (figure 4 (a), (b) and (c) panels).

**Conclusions**

We have studied the vortex dynamics in a periodic potential created by local enhancement of superconductivity. This has been achieved by an array of Nb nanotriangles embedded in a V film of slightly lower critical temperature. In addition, the nanodot area and the area between nanodots are very similar. The most remarkable findings are the following: i) the Nb nanotriangles act as antipinning defects, quenching the main commensurability effect between the vortex lattice and the defect (Nb dots) unit cell. ii) Hysteresis effect in the magnetoresistance appears when the number of vortices increases and the vortex lattice begins to probe the Nb nanotriangles. The magnetoresistance hysteresis is still present when commensurability effects are washed out by decreasing the temperature. iii) We observe a broadening of the vortex liquid phase and a softening of the vortex lattice (decreasing pinning force) in comparison with similar hybrid systems; i. e. superconducting films with embedded nanotriangles of non-superconducting defects.

In summary, the local enhancement of the superconductivity created by the Nb nanotriangles gives rise to new outcomes in the vortex dynamics that can be modified by external parameters such as temperature and applied magnetic fields parallel to the sample plane.

*Acknowledgments*
We want to thank I. K. Schuller and V. Bekeris for illuminating discussions, M. Buchacek for useful suggestions and J. Romero (CAI Tecnicas Fisicas, UCM) for his help with the SQUID measurements. The work has been supported by Spanish MICINN grants FIS2016-76058 (AEI/FEDER, UE), EU COST- CA16218. IMDEA Nanociencia acknowledges support from the 'Severo Ochoa' Programme for Centres of Excellence in R&D (MICINN, Grant SEV-2016-0686). AG acknowledges financial support from Spanish MICINN Grant ESP2017-86582-C4-1-R and IJCI-2017-33991. JdV thanks Fundacion Ramon Areces for a postdoctoral fellowship.

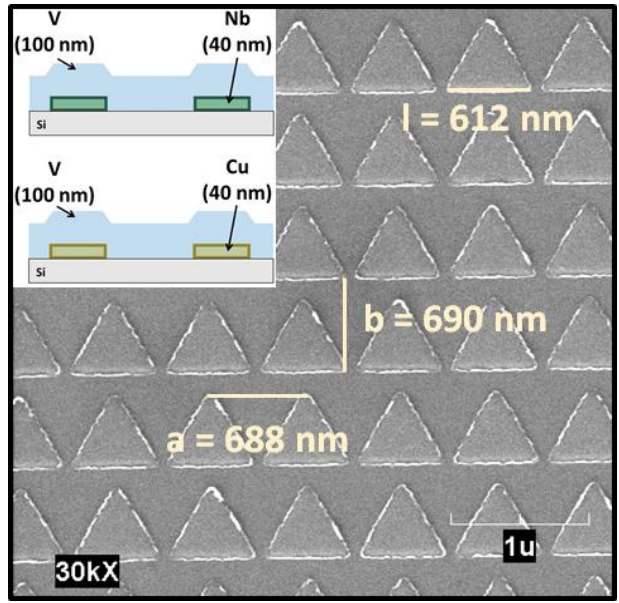

**Figure 1**. Scanning Electron Microscope image of the array of nanotriangles. The dimension, and periodicity of the nanotriangles are shown in the image. Inset shows sketch of both MS sample (upper drawing) and WS sample (lower drawing), not to scale. Bar scale at the bottom right corner is 1 μm.

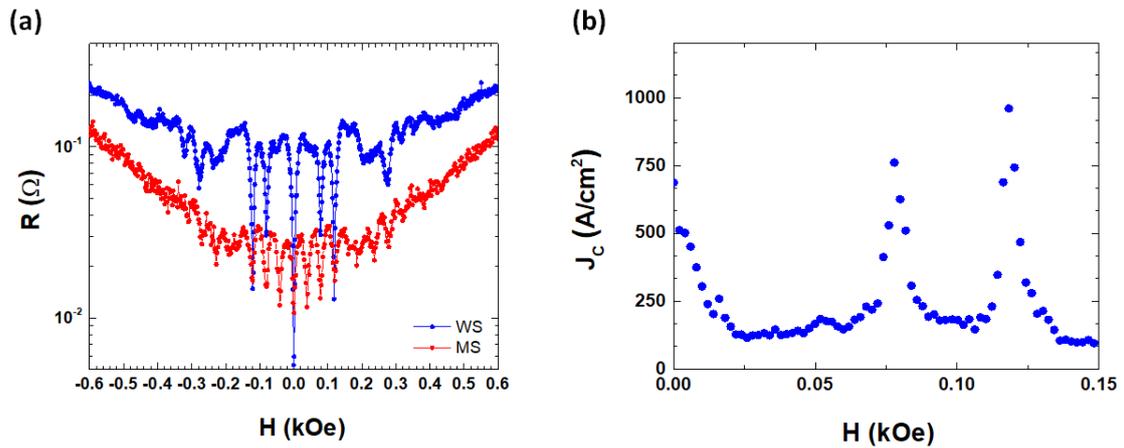

**Figure 2.** MS and WS samples at 0.97 $T_{c0}$. $T_{c0} = T_c$ (H=0) (a) Y-axis: Resistance; X-axis: Applied magnetic field; WS sample red plot, MS sample blue plot; (b) MS sample Y – axis: Critical current density. X – axis: Applied magnetic field.

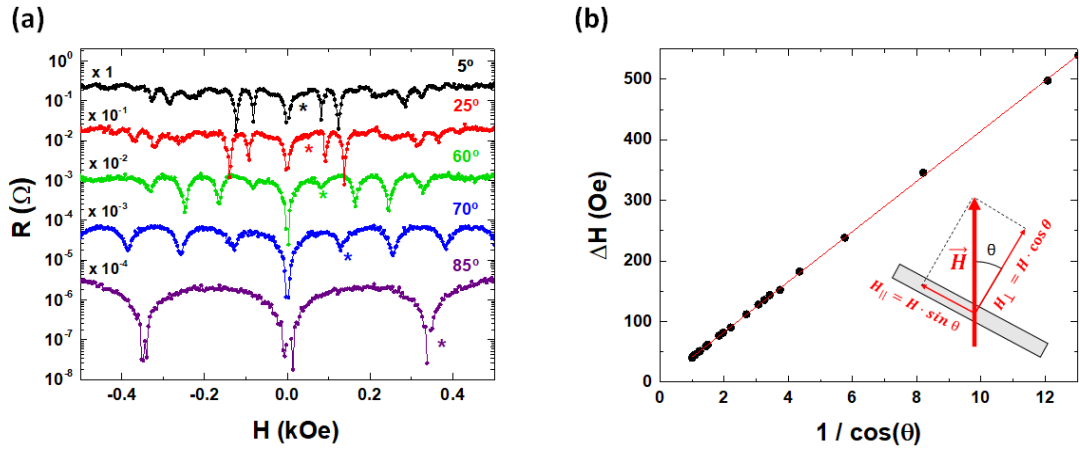

**Figure 3.** (a) Sample MS: Resistivity vs applied magnetic field curves for different angles θ between the field and the direction perpendicular to the sample plane. The stars mark the position of the first matching field. T = 0.98 $T_{c0}$. The experimental plots have been vertically displaced. (b) Sample MS: the angular dependence of the distance between consecutive minima ΔH (0) = 39 Oe. The solid line is a fit to the expression ΔH (θ) = ΔH (0) / cos (θ). Inset shows a sketch of the experimental geometry.

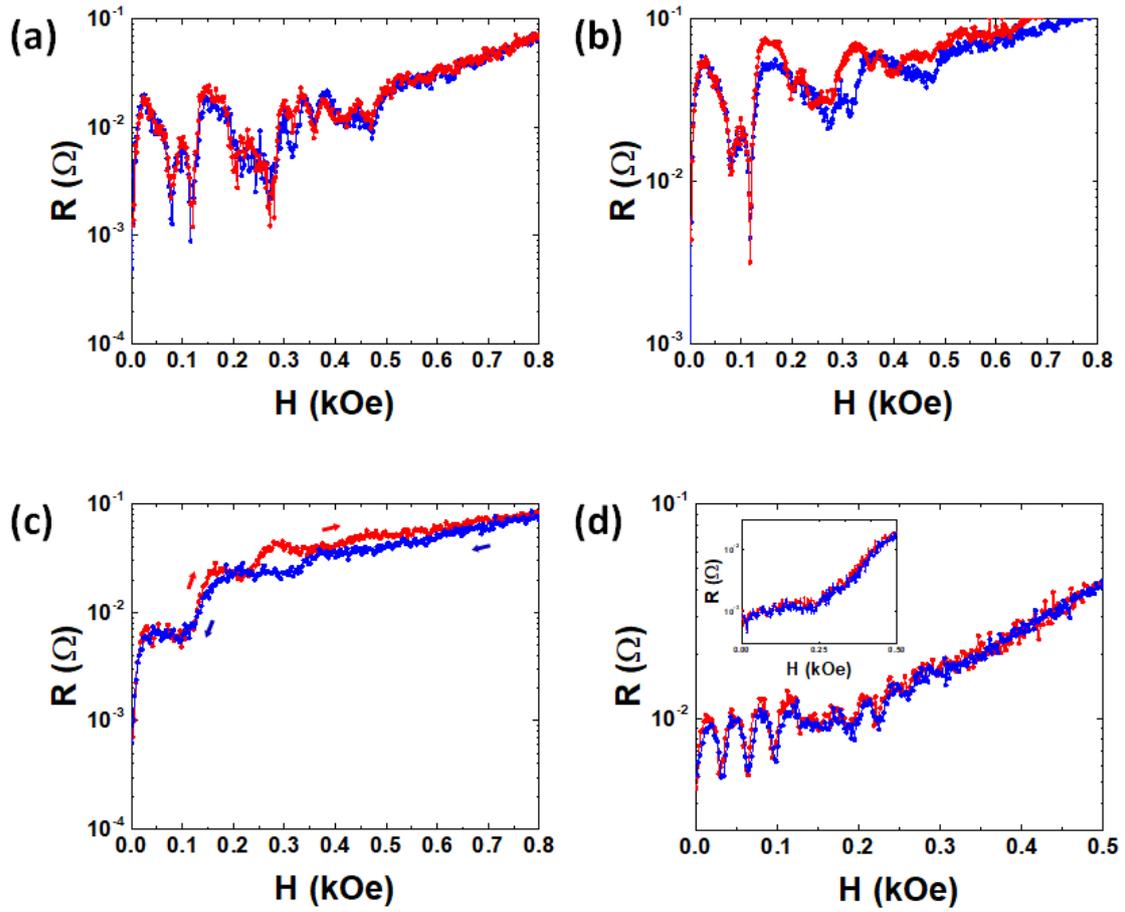

**Figure 4.** Resistance versus applied magnetic fields. Red curves are measured increasing the magnetic fields and blue curves are measured decreasing the magnetic fields, as indicated in panel (c) with red arrows and blue arrows respectively. Sample MS: Panels (a) T = 0.94 $T_{c0}$; (b) T = 0.92 $T_{c0}$ and (c) T = 0.80 $T_{c0}$. Sample WS: Panel (d) T = 0.97 $T_{c0}$; inset T = 0.90 $T_{c0}$. $T_{c0}$ = $T_c$ (H=0).

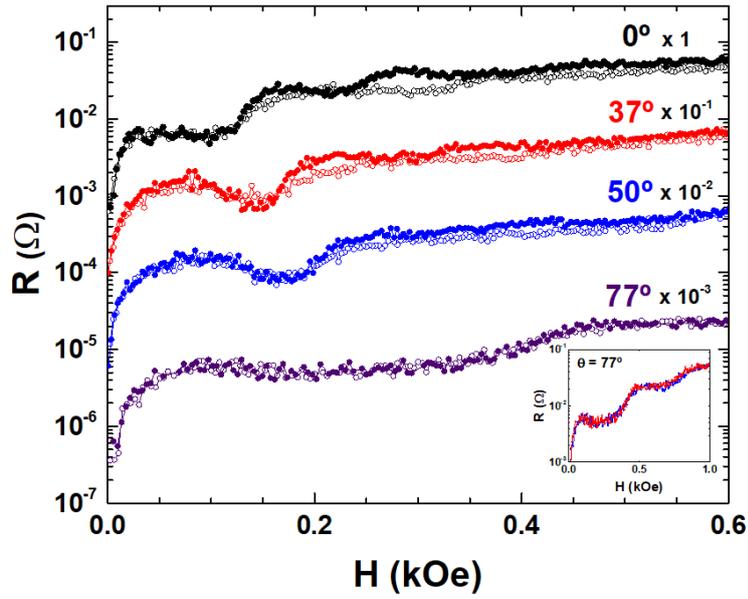

**Figure 5.** Sample MS at T = 0.80 T$_{c0}$. Resistance versus magnetic applied fields at different angles between the normal to the sample plane and the direction of the applied fields, for the experiment geometry see inset figure 3(b). Increasing applied magnetic field full symbols and decreasing applied magnetic fields empty symbols. Inset shows the magnetoresistance up to 1 kOe and with tilted angle θ =77 °.

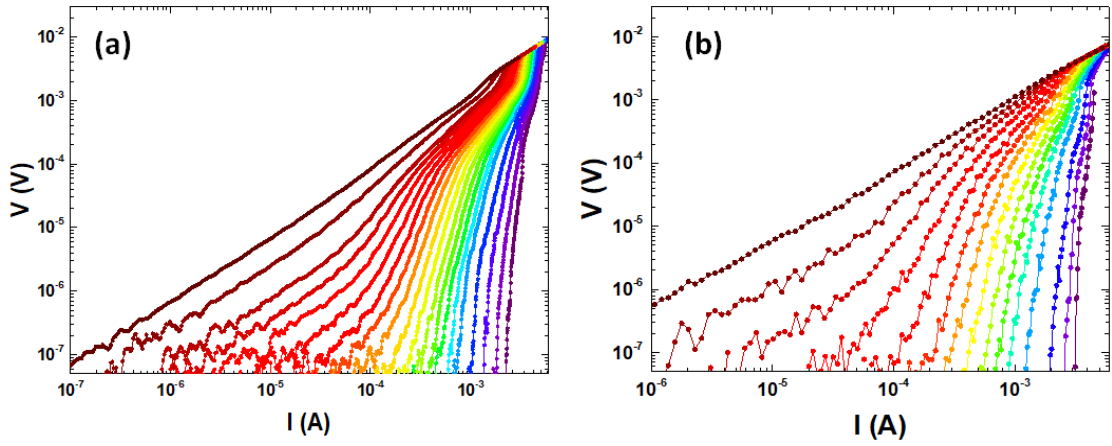

**Figure 6.** (a) I-V isotherms from 0. 89 $T_{c0}$ to 0.99 $T_{c0}$ for sample MS ($T_{c0}$ = 4.25 K). Data taken every 20 mK (applied magnetic field H= 117 Oe); (b) isotherms from T = 0.91 $T_{c0}$ to T = 0.99 $T_{c0}$ for sample WS ($T_{c0}$ = 4.38 K). Data taken every 20 mK between 0.99 $T_{c0}$ and 0.94 $T_{c0}$ and 40 mK between 0.94 $T_{c0}$ and 0.91 $T_{c0}$ (applied magnetic field H= 117 Oe).

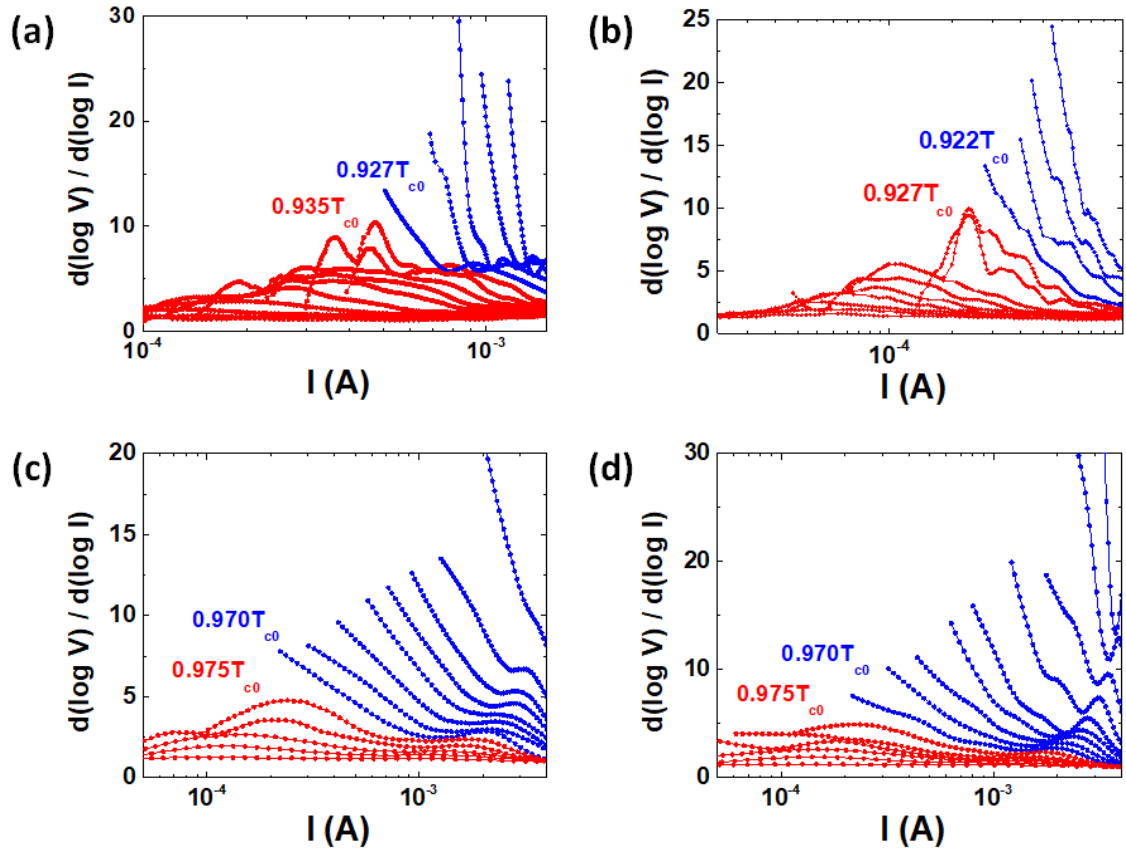

**Figure 7.** Derivatives of the log (V) - log (I) curves as a function of the current. Sample MS ($T_{c0}$ = 4.25 K) with applied magnetic fields: (a) H= 117 Oe and (b) H= 273 Oe. Sample WS ($T_{c0}$ = 4.38 K) with applied magnetic fields: (c) H= 117 Oe and (d) H= 273 Oe.

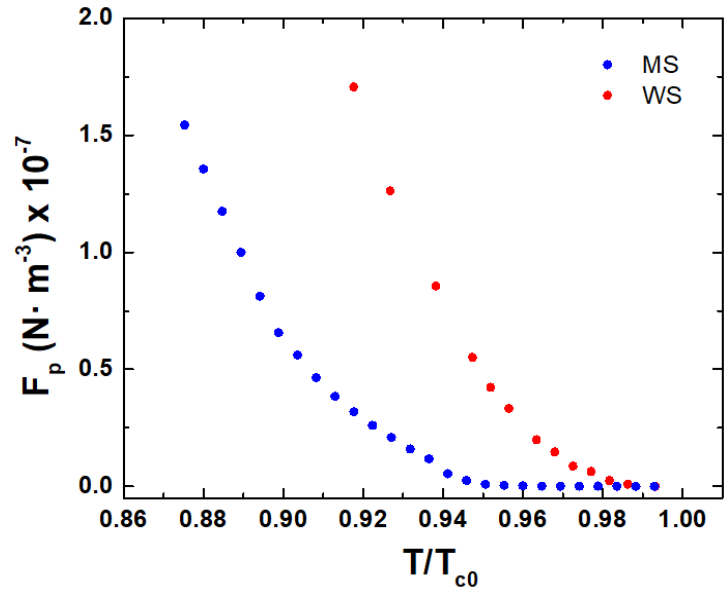

**Figure 8.** Y-axis pinning force ($F_p$ = $J_c$ B; $J_c$ and B being the critical current and the magnetic field respectively); X-axis normalized critical temperature, $T_{c0}$ = $T_c$ (H=0). At H = 39 Oe; filled symbol sample MS, open symbol sample WS.